\documentclass[twocolumn,showpacs,preprintnumbers,amsmath,amssymb]{revtex4}

\usepackage{graphicx}
\usepackage{dcolumn}
\usepackage{bm}

\begin{document}

\preprint{}

\title{Measurements of the density-dependent many-body electron mass \\in 2D GaAs/AlGaAs Heterostructures
}

\author{Y.-W. Tan$^{1}$}
\author{J. Zhu$^{2}$}
\author{H. L. Stormer$^{1,3,4}$}
\author{L. N. Pfeiffer$^{4}$}
\author{K. W. Baldwin$^{4}$}
\author{K. W. West$^{4}$}

\affiliation{$^{1}$Department of Physics, Columbia University, New
York, New York 10027\\ $^{2}$Department of Physics, Cornell
University, Ithaca, New York 14853\\
$^{3}$Department of Applied Physics and Applied Mathematics,
Columbia University, New York, New York 10027\\ $^{4}$Bell Labs,
Lucent Technologies, Murray Hill, New Jersey 07974
}%

\date{November 24, 2004}

\begin{abstract}
We determine the density-dependent electron mass, m*, in
two-dimensional (2D)
 electron systems of GaAs/AlGaAs heterostructures by performing detailed
 low-temperature Shubnikov deHaas measurements. Using very high quality
 transistors with tunable electron densities we measure m* in single,
 high mobility specimens over a wide range of $r_{s}$ (6 to 0.8). Toward
 low-densities we observe a rapid increase of m* by as much as 40\%. For $2>r_{s}>0.8$
 the mass values fall $\sim10\%$ \textit{below} the band mass of GaAs. Numerical calculations are
 in qualitative agreement with our data but differ considerably in detail.
\end{abstract}

\maketitle

In a crystal the mass of an electron often differs from its mass,
$m_{0}$, in free space. The electron mass in a semiconductor can
deviate from $m_{0}$ by more than a factor of ten. The origin of
this effect is interference of the electron wavefunction with the
periodic array of the ions in the solid. Such ``single particle"
effects are well understood and readily calculated. However, there
are other factors that affect the electron mass. In general, any
excitation of the solid -- such as phonons, spin waves, plasmons
-- can impact the dispersion relation of the carrier\cite{Pines},
but only if such excitations come close to resonance. More
intricate yet, the electron mass is modified by interactions with
all neighboring conduction electrons. Naively one would think that
interactions always enhance the carrier mass since they imply
pushing against other electrons, making them apparently heavier.
Yet, curiously, theory tells us that the mass can be reduced as
well\cite{Rice}. Such interactions are of complex ``many particle"
origin and bring us to the edge of the theoretical and numerical
abilities in condensed matter theory.

The impact of electron-electron (e-e) interaction on carrier mass
increases on lowering the dimensionality. Three dimensional (3D)
electron systems are expected to show little effect, whereas 1D
systems are highly affected. From an experimental point of view,
3D systems are most abundant and best characterized, whereas 1D
systems are rare and suffer from many complications such as
Peierls instabilities and their sensitivity to disorder.
Two-dimensional electron systems (2DESs) provide an excellent
compromise for the study of such many-body phenomena: 2DESs have
been honed to extremely high quality\cite{MBE} and the expected
effects are moderately strong. In addition, the electron density
in a 2DES is in principle continuously tunable, allowing the study
of such phenomena within a single specimen over a wide range of
densities. The strength of e-e interaction is usually described by
a dimensionless parameter $r_{s}$, which is defined as the ratio
of Coulomb interaction to Fermi energy. In 2D, $r_{s}$ is
inversely proportional to the square root of density, so a
variable density translates into a tunable e-e interaction.

Starting in the late sixties\cite{Rice}, there have been many
theoretical studies\cite{SMG} of the impact of e-e interaction on
carrier mass using ever more powerful numerical tools.
Experimentally, the Silicon
Metal-Oxide-Semiconductor-Field-Effect-Transistors (MOSFETs) had
been the dominant implementation of a 2DES. Smith and
Stiles\cite{Smith&Stiles} were the first to measure m* as a
function of $r_{s}$ in such a device. However, subsequent work by
Fang et al.\cite{FangFowler} asserted that such mass measurements
in MOSFETs were affected by several side effects. Yet, the Smith
and Stiles\cite{Smith&Stiles} data remain the experimental
reference point for this rapidly progressing area of theoretical
investigation.

Recently, Coleridge et al.\cite{Coleridge} have performed mass
measurements on five fixed density GaAs/AlGaAs heterostructures,
but only for $r_{s}<1.7$. They observed a monotonically increasing
m* with increasing carrier density. With the recent interest in a
transition from an electron liquid to an electron solid at very
high $r_{s}$, several new studies have taken
place\cite{Gershenson,Kravchenko}. They concentrate on phenomena
associated with the anticipated phase transition but not with the
electron liquid \textit{per se}. Given the theoretical progress in
the area of e-e interactions and the availability of very
high-quality 2DESs, a careful measurement of the effective 2D
electron mass over a wide range of density seems to be of
considerable importance to make contact between theory and
experiment.

Toward this end we have measured m* in a very high quality,
tunable density, GaAs/AlGaAs heterojunction-insulated gate field
effect transistor (HIGFET) for $6>r_{s}>0.8$. We observe a
strongly increasing mass towards low densities and a mass
$\sim10\%$ below the band mass for all $2>r_{s}>0.8$.

Our primary sample, HIGFET-1, was grown by molecular beam epitaxy
(MBE) onto a $(001)$ GaAs substrate. The 2DES resides at the
interface of 5$\mu$m of GaAs and 5nm of AlAs, topped with 4$\mu$m
of Al$_{0.33}$Ga$_{0.67}$As. The latter acts as an insulator
separating the channel from the gate, which consists of a $25nm$
thick heavily doped, conducting GaAs $n^{+}$ layer. The material
was processed into a 600$\mu$m square mesa using photolithography
and contacted via Ni-Ge-Au annealed pads. An extra pad contacted
the gate. Another sample, HIGFET-2, differs from HIGFET-1 in a
thinner channel material (2$\mu$m GaAs) and the lack of the thin
AlAs layer. Both samples have a peak mobility of
$\sim1\times10^{7}cm^{2}/Vs$. The electron density of the 2DEG can
be tuned by changing the gate voltage of the transistor. This
provides a major advantage as compared to fixed density specimens
since the strength of e-e interaction can be changed continuously.

Measurement of the specific heat would be the most direct way to
determine the mass that includes e-e interactions. Such
experiments are exceedingly challenging and have not yet been
realized with high precision\cite{heatCv}. A mass determined by
cyclotron resonance will not exhibit effects from e-e interactions
according to Kohn's theorem\cite{Kohn}. Instead, we employ the
Shubnikov deHaas (SdH) effect and determine the effective mass
from the reduction of the amplitude of these magnetoresistance
oscillations with increasing temperature. For SdH measurements to
provide reliable mass data, data collection has to be performed
within appropriate parameter windows, and the interpretation of
the data has to be conducted with considerable care and multiple
cross checks. In the following paragraphs we detail our procedure.

\begin{figure}
\includegraphics[width=90mm]{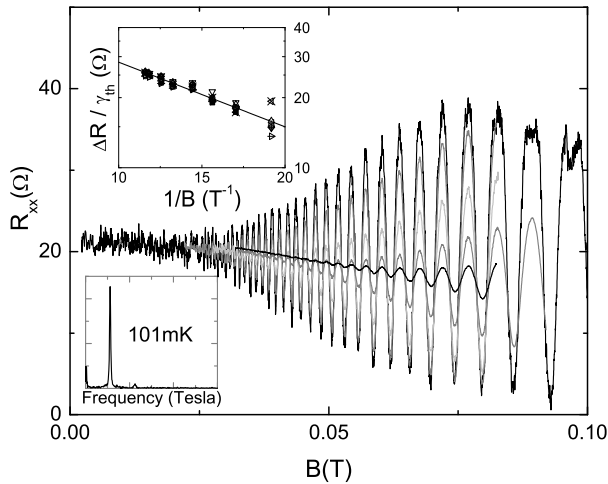}
\caption{\label{fig:SdH} Shubnikov deHaas (SdH) oscillations for
$n=5.4\times10^{10}/cm^{2}$ at T=101, 180, 260, 380mK. The lower
inset: Fourier spectrum from the 101mK data vs 1/B. Upper inset:
all Dingle plots for seven different T ranging from 100 to 400mK
data sets yield similar scattering time $\tau_{q}$s.}
\end{figure}

All measurements were performed in a dilution refrigerator over a
temperature range from 100 to 400mK. We used conventional low
frequency AC lock-in techniques with excitation currents ranging
from 1nA to 100nA, chosen to avoid sample heating. Data were taken
in a single cool down in order to have the most consistent quality
from each sample. The temperature is based on a calibrated
Ruthenium oxide thermometer mounted on the same silver sample
holder. Magneto-resistance is negligible at low fields and the
relatively high temperature ensures good thermal coupling between
sample and thermometer. At each density, set by the gate voltage,
we recorded a family of SdH traces at a range of temperatures.
Data collection was limited to the moderate magnetic field region
such that the SdH oscillations were well developed but before the
small spin splitting in GaAs was resolved. Therefore, in the
region of our measurements, each minimum in the oscillations
corresponds to a Landau level index $i$. The temperature was kept
sufficiently high to avoid the quantum Hall regime, in which SdH
oscillations are becoming non-sinusoidal.

SdH oscillations are a result of the comb of Landau levels
sweeping through the Fermi level while the magnetic field is
ramped. Hence, the oscillating part of the magneto resistivity
$\Delta\rho_{xx}$ can be written as a Fourier
series:\cite{ShoenbergBK,Igor_B},
\begin{equation}
\Delta\rho_{xx}(\varepsilon)=\rho_{0}\sum_{p}\gamma_{th}
c_{\alpha,p}\exp(-\frac{p\pi}{\omega_{c}^{*}\tau_{q}})\cos[2\pi
p(\frac{\varepsilon}{\hbar\omega_{c}^{*}}-\frac{1}{2})],
\end{equation}
with $\omega_{c}^{*}=eB/m^{*}$, and $c_{\alpha,p}$ being a
disorder coefficient\cite{Igor_B}. The T-dependence $\gamma_{th}$
is given by
\begin{equation}
\gamma_{th}=\frac{p\cdot2\pi^{2}k_{B}T/\hbar\omega_{c}^{*}}
{\sinh(p\cdot2\pi^{2}k_{B}T/\hbar\omega_{c}^{*})}
\end{equation}
It is common practice to maintain only the fundamental term and
neglect all higher order Fourier components. We will later
explicitly check this assumption for our measurements. We derive
the effective electron mass, m*, by fitting, on a $\log(\Delta
R/T)$ vs $T$ plot, expression (2) to our T-dependent data from
each index $i$, where $R$ is resistance. We achieve excellent fits
to all SdH data with a single mass value for each density in the
whole temperature range $100mK<T<400mK$\cite{YZhang}. Before
discussing these results we performed several cross-checks to
establish the reliability of our data reduction.

In recent literature there arose a concern as to whether the 3D
SdH formalism applies correctly to 2D cases\cite{2DLK3}.
Significant deviations from the original Lifshitz and Kosevich
(LK) formula were observed in de Haas-van Alphen effects of
layered organic conductors\cite{dHvA1} as well as in 2DESs of
GaAs\cite{dHvA2}. Such deviations can be traced back to
significant contribution from higher order Fourier components.
When higher harmonics are negligible, the thermal reduction of the
amplitude is well described by eq.(2)\cite{2DLKth1,2DLKth2}.

In order to explore the variability of our mass data due to such
possible deformations, we have used three different methods to
derive m*. i) Data points depicted by solid black circles in
fig.2(a) are m* values fitted directly to eq.(2) with p=1 only.
Each point represents the average mass from several different
Landau level indices, and this root-mean-square deviation
dominates the error bars. ii) Masses depicted as solid squares are
derived from the same data, but after Fourier filtering them on a
1/B plot. A typical Fourier spectrum is shown in the inset of
fig.1, demonstrating that higher harmonics are practically
negligible. Accordingly, the filtered and unfiltered data differ
only slightly from one another. iii) We follow a revised version
of the LK derivation for 2D Fermi liquids under strong e-e
interaction and electron-impurity scattering\cite{Maslov}. There,
the thermal reduction of the first Fourier component of the SdH
amplitude follows an exponential decay form instead of eq.(2). The
masses obtained from these fits are shown as crosses in fig.2(a).
These m* values are close to those from the Fourier filtered
method. We conclude that our SdH data fall within the window over
which an LK analysis can be applied and that our mass derivation
is robust.

\begin{figure}
\includegraphics[width=90mm]{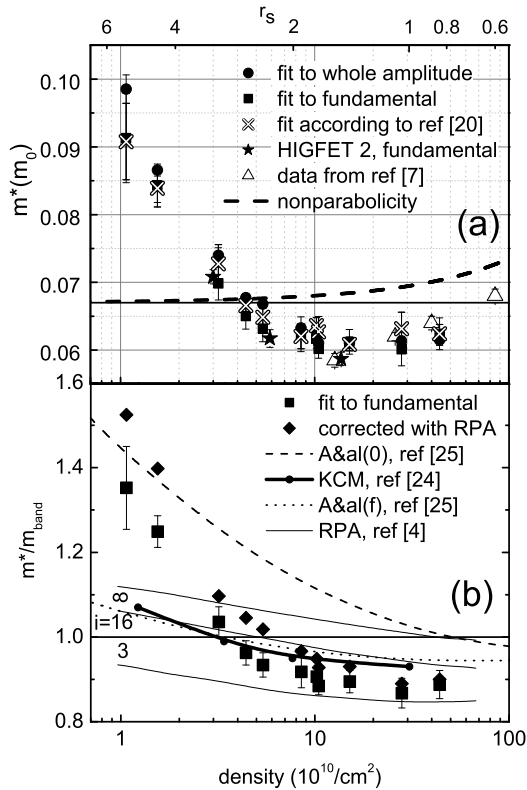}
\caption{\label{fig:mass} Density dependence of 2D electron
effective mass m* (a) m* from three different fitting methods,
from a second specimen, and from another group\cite{Coleridge},
see legend. Mass enhancement due to non-parabolicity shown as
dashed line. (b) m*s of ``fundamental fit" in (a) corrected for
nonparabolicity. Several theories are indicated as lines, see
legend and text.}
\end{figure}

Another possible source of concern are density inhomogeneities.
Such inhomogeneities were shown to considerably affect the
determination of the scattering time via SdH\cite{inhomogeneity};
a subject we will address later in this paper. From the known
density gradient across the wafer, we determine the density
gradient to be $\lesssim0.07\%$ across our sample. The
contribution of density inhomogeneities from surface roughness at
the interface is estimated to be $\lesssim0.1\%$\cite{MBEGaAs}.
Our analytical calculations and many numerical simulations
demonstrate that the thermal reduction factor is not affected by
such density inhomogeneity and the m* readings at any fixed field
will not be influenced.

Data from HIGFET-2 are shown as stars in fig.2(a). Despite the
differences between the two samples, their m* coincide within the
error bars. We also include data from four fixed density samples
measured earlier by Coleridge et al.\cite{Coleridge} at higher
densities. These data are also consistent with our results. The m*
data from Si-MOSFETs are generally $\sim20\%$ larger than GaAs
data.\cite{Smith&Stiles,Gershenson,Kravchenko}

The effective mass versus density data of fig.2(a) follows a
smooth but non-monotonic curve. With increasing $r_{s}$ at the low
density part, there is a strong enhancement of m*. At
$r_{s}\sim5$, m* exceeds the GaAs band mass by $\sim40\%$. In the
high density region, m* stays $\sim10\%$ below the band mass for
$2\geq r_{s}\geq0.8$. In this regime the
non-parabolicity\cite{nonparabolicity} of the GaAs conduction band
actually further enhances the discrepancy between single particle
mass and our measured mass. The dashed line shows the result of
numerical calculation for the band mass at the Fermi energy due to
2D confinement plus band filling. Results from three different
trial wave functions fall within the thickness of the line. The
ratio of the measured m* to the band mass, corrected for this
non-parabolicity, are shown in fig.2(b) together with theoretical
calculations. For clarity, we limit the data points to a subset of
the data of fig.2(a).

Kwon, Ceperley, and Martin\cite{KCM}(KCM) using a variational
Monte Carlo method, and Asgari et al.\cite{Polini}(A\&al) using a
many-body local field approach have performed extensive
theoretical calculations of the impact of e-e interactions on the
mass in 2D. The results of these numerics are plotted in fig.2(b)
as a full solid line and as a dashed line, respectively. The KCM
theory reproduces quite well the average mass values at low
$r_{s}$, whereas the A\&al(0) theory seems to depict the overall
shape and the upturn at high $r_{s}$. However, both theories
assume a zero-thickness 2DES. When adjusted for finite thickness,
the A\&al result follows the dotted line, A\&al(f)\cite{Polini}.
No such adjustment is available for KCM and the coincidence of
both curves must be considered accidental. In any case, these
calculations seem to capture some aspects of the data, but clearly
do not describe the high $r_{s}$ regime.

All numerical calculations have been performed for a 2D system in
the absence of a magnetic-field. Since our data were collected in
small B-fields and mostly in very high Landau levels we can regard
our data as representing this limiting $B=0$ case. However, Smith,
MacDonald and Gumbs\cite{SMG} (SMG) have performed mass
calculations in the presence of a B-field based on the
random-phase approximation (RPA). Their results are shown as a
sequence of thin lines identified by the representative Landau
index $i$.  We note that $i=\infty$ is equivalent to $B=0$ and can
be directly compared to the other calculations, showing further
the discrepancies between different theories. These $i=\infty$
results also differ considerably from our data. At the same time,
the SMG calculation shows a considerable dependence of the mass
value on the index which needs to be taken into account.

The experimental window dictated by temperature, sinusoidal
lineshape and available B-field, results in a correlation between
density n and available index $i$. Lower densities require the
recording of low indices whereas higher densities allow to measure
much higher indices. While the $i=\infty$ SMG results differ
considerable from our data, it appears reasonable to use their
fractional dependence of the mass on index,
$m^{*}(i=\infty)/m^{*}(i)$, to correct our data for such an index
dependence. The resulting m* data are shown as filled diamonds in
fig.2(b). The overall shape of the density dependent mass is not
much affected although the very low density masses are enhanced
beyond the previous error bars, since such data were taken at
relatively low $i$. On the other hand the high density data are
almost unaffected. While the correction suggested by the SMG work
modifies somewhat the comparison between theory and experiment,
the overall conclusions remain intact.

In addition to m*, the envelope of the SdH oscillations provides
us with a measure of the quantum scattering time $\tau_{q}$ at
each temperature. Since $\tau_{q}$ enters the LK expression,
eq.(1), a temperature dependence of $\tau_{q}$ could affect the
value of m*. Before examining the data, we should stress that a
T-dependence of $\tau_{q}$ must be considered very weak and only
on the scale of $T/T_{F}$, $T_{F}$ being the Fermi temperature,
since scattering by fixed imperfections is the only remaining
mechanism and it is practically T-independent in our temperature
and density range. Nevertheless, we evaluated $\tau_{q}$ employing
the semi-log Dingle plot of the SdH amplitude normalized to
$\rho_{0}\gamma_{th}$(eq.(1)) vs $1/B$. The upper inset of fig.1
shows such a typical Dingle plot for seven different temperatures.
The slope of the data determines $\tau_{q}$, which can be affected
by density inhomogeneities\cite{inhomogeneity}. However, here we
pursue only a possible T-dependence of $\tau_{q}$. From extensive
modeling we determine that it is unaffected by inhomogeneity of as
much as 10\%. Evaluating many Dingle plots, we find that over our
T-range, $\tau_{q}(T)$ varies by less than $1\%$ for all $r_s<4$,
see also ref.\cite{Gershenson}. At the lowest two densities, our
Dingle plots are ill-defined. From numerical simulations we deduce
that even there a T-dependence of $\tau_{q}$ on the scale of
$T/T_{F}$ will at most generate a 3\% error in m*. Therefore, the
effect of a T-dependent $\tau_{q}$ will have an insignificant
impact on m* in fig.2(a).

In conclusion, we have performed high precision measurements of
the electronic effective mass in an ultra-high quality, tunable,
two-dimensional electron system over a wide rage of $r_{s}$,
$6>r_{s}>0.8$. Performing various cross checks and exploring
several sources for error we convinced ourselves that our data
provide an accurate measurement of the density dependent impact of
electron-electron interactions on the electron mass in a 2D
system. Over wide stretches of density this mass renormalization
can be negative. Various theoretical calculations reproduce
sections of our data quite well but none shows good agreement for
the whole range of densities.

We thank I. Aleiner, A. Millis, and A. Mitra for discussions, and
Asgari et al. for clarification on Ref. \onlinecite{Polini}. This
work is supported by NSF under DMR-03-52738, by the DOE under
DE-AIO2-04ER46133, and by a grant of the W. M. Keck Foundation.


\begin{thebibliography}{26}
\expandafter\ifx\csname
natexlab\endcsname\relax\def\natexlab#1{#1}\fi
\expandafter\ifx\csname bibnamefont\endcsname\relax
  \def\bibnamefont#1{#1}\fi
\expandafter\ifx\csname bibfnamefont\endcsname\relax
  \def\bibfnamefont#1{#1}\fi
\expandafter\ifx\csname citenamefont\endcsname\relax
  \def\citenamefont#1{#1}\fi
\expandafter\ifx\csname url\endcsname\relax
  \def\url#1{\texttt{#1}}\fi
\expandafter\ifx\csname
urlprefix\endcsname\relax\def\urlprefix{URL }\fi
\providecommand{\bibinfo}[2]{#2}
\providecommand{\eprint}[2][]{\url{#2}}

\bibitem[{\citenamefont{Pines}(1963)}]{Pines}
\bibinfo{author}{\bibfnamefont{D.}~\bibnamefont{Pines}},
  \emph{\bibinfo{title}{Elementary Excitations in Solids}}
  (\bibinfo{publisher}{W. A. Benjamin, Inc.}, \bibinfo{address}{New York},
  \bibinfo{year}{1963}).

\bibitem[{\citenamefont{Rice}(1965)}]{Rice}
\bibinfo{author}{\bibfnamefont{T.~M.} \bibnamefont{Rice}},
  \bibinfo{journal}{Annl.\ Phys.} \textbf{\bibinfo{volume}{31}},
  \bibinfo{pages}{100} (\bibinfo{year}{1965}).

\bibitem[{\citenamefont{Pfeiffer and West}(2003)}]{MBE}
\bibinfo{author}{\bibfnamefont{L.}~\bibnamefont{Pfeiffer}} \bibnamefont{and}
  \bibinfo{author}{\bibfnamefont{K.~W.} \bibnamefont{West}},
  \bibinfo{journal}{Physica E} \textbf{\bibinfo{volume}{20}},
  \bibinfo{pages}{57} (\bibinfo{year}{2003}).

\bibitem[{\citenamefont{Smith et~al.}(1992)\citenamefont{Smith, MacDonald, and
  Gumbs}}]{SMG}
\bibinfo{author}{\bibfnamefont{A.~P.}~\bibnamefont{Smith}},
  \bibinfo{author}{\bibfnamefont{A.~H.}~\bibnamefont{MacDonald}},
  \bibnamefont{and} \bibinfo{author}{\bibfnamefont{G.}~\bibnamefont{Gumbs}},
  \bibinfo{journal}{Phys. Rev. B} \textbf{\bibinfo{volume}{45}},
  \bibinfo{pages}{R8829} (\bibinfo{year}{1992}), \bibinfo{note}{and references
  therein}.

\bibitem[{\citenamefont{Smith and Stiles}(1972)}]{Smith&Stiles}
\bibinfo{author}{\bibfnamefont{J.~L.} \bibnamefont{Smith}} \bibnamefont{and}
  \bibinfo{author}{\bibfnamefont{P.~J.} \bibnamefont{Stiles}},
  \bibinfo{journal}{Phys. Rev. Lett.} \textbf{\bibinfo{volume}{29}},
  \bibinfo{pages}{102} (\bibinfo{year}{1972}).

\bibitem[{\citenamefont{Fang et~al.}(1977)\citenamefont{Fang, Fowler, and
  Hartstein}}]{FangFowler}
\bibinfo{author}{\bibfnamefont{F.~F.} \bibnamefont{Fang}},
  \bibinfo{author}{\bibfnamefont{A.~B.} \bibnamefont{Fowler}},
  \bibnamefont{and}
  \bibinfo{author}{\bibfnamefont{A.}~\bibnamefont{Hartstein}},
  \bibinfo{journal}{Phys. Rev. B} \textbf{\bibinfo{volume}{16}},
  \bibinfo{pages}{4446} (\bibinfo{year}{1977}).

\bibitem[{\citenamefont{Coleridge et~al.}(1996)}]{Coleridge}
\bibinfo{author}{\bibfnamefont{P.~T.} \bibnamefont{Coleridge}}
  \bibnamefont{et~al.}, \bibinfo{journal}{Surf. Sci.}
  \textbf{\bibinfo{volume}{361/362}}, \bibinfo{pages}{560}
  (\bibinfo{year}{1996}).

\bibitem[{\citenamefont{Pudalov et~al.}(2002)}]{Gershenson}
\bibinfo{author}{\bibfnamefont{V.~M.} \bibnamefont{Pudalov}}
  \bibnamefont{et~al.}, \bibinfo{journal}{Phys. Rev. Lett.}
  \textbf{\bibinfo{volume}{88}}, \bibinfo{pages}{196404}
  (\bibinfo{year}{2002}).

\bibitem[{\citenamefont{Shashkin et~al.}(2003)}]{Kravchenko}
\bibinfo{author}{\bibfnamefont{A.~A.} \bibnamefont{Shashkin}}
  \bibnamefont{et~al.}, \bibinfo{journal}{Phys. Rev. Lett.}
  \textbf{\bibinfo{volume}{91}}, \bibinfo{pages}{046403}
  (\bibinfo{year}{2003}).

\bibitem[{\citenamefont{Bayot et~al.}(1996)}]{heatCv}
\bibinfo{author}{\bibfnamefont{V.}~\bibnamefont{Bayot}} \bibnamefont{et~al.},
  \bibinfo{journal}{Phys. Rev. Lett.} \textbf{\bibinfo{volume}{76}},
  \bibinfo{pages}{4584} (\bibinfo{year}{1996}).

\bibitem[{\citenamefont{Kohn}(1961)}]{Kohn}
\bibinfo{author}{\bibfnamefont{W.}~\bibnamefont{Kohn}}, \bibinfo{journal}{Phys.
  Rev.} \textbf{\bibinfo{volume}{123}}, \bibinfo{pages}{1242}
  (\bibinfo{year}{1961}).

\bibitem[{\citenamefont{Shoenberg}(1984)}]{ShoenbergBK}
\bibinfo{author}{\bibfnamefont{D.}~\bibnamefont{Shoenberg}},
  \emph{\bibinfo{title}{Magnetic Oscillations in Metals}}
  (\bibinfo{publisher}{Cambridge University Press},
  \bibinfo{address}{Cambridge}, \bibinfo{year}{1984}).

\bibitem[{\citenamefont{Vavilov and Aleiner}(2004)}]{Igor_B}
\bibinfo{author}{\bibfnamefont{M.~G.} \bibnamefont{Vavilov}} \bibnamefont{and}
  \bibinfo{author}{\bibfnamefont{I.~L.} \bibnamefont{Aleiner}},
  \bibinfo{journal}{Phys. Rev. B} \textbf{\bibinfo{volume}{69}},
  \bibinfo{pages}{035303} (\bibinfo{year}{2004}).

\bibitem[{\citenamefont{DasSarma et~al.}(2004)\citenamefont{DasSarma, Galitski,
  and Zhang}}]{YZhang}
\bibinfo{author}{\bibfnamefont{S.}~\bibnamefont{Das Sarma}},
  \bibinfo{author}{\bibfnamefont{V.~M.} \bibnamefont{Galitski}},
  \bibnamefont{and} \bibinfo{author}{\bibfnamefont{Y.}~\bibnamefont{Zhang}},
  \bibinfo{journal}{Phys. Rev. B} \textbf{\bibinfo{volume}{69}},
  \bibinfo{pages}{125334} (\bibinfo{year}{2004}).

\bibitem[{\citenamefont{Champel}(2001)}]{2DLK3}
\bibinfo{author}{\bibfnamefont{T.}~\bibnamefont{Champel}},
  \bibinfo{journal}{Phys. Rev. B} \textbf{\bibinfo{volume}{64}},
  \bibinfo{pages}{054407} (\bibinfo{year}{2001}), \bibinfo{note}{and references
  therein.}

\bibitem[{\citenamefont{Harrison et~al.}(1996)}]{dHvA1}
\bibinfo{author}{\bibfnamefont{N.}~\bibnamefont{Harrison}}
  \bibnamefont{et~al.}, \bibinfo{journal}{Phys. Rev. B}
  \textbf{\bibinfo{volume}{54}}, \bibinfo{pages}{9977} (\bibinfo{year}{1996}),
  \bibinfo{note}{and references therein.}

\bibitem[{\citenamefont{Wiegers et~al.}(1997)}]{dHvA2}
\bibinfo{author}{\bibfnamefont{S.}~\bibnamefont{Wiegers}} \bibnamefont{et~al.},
  \bibinfo{journal}{Phys. Rev. Lett.} \textbf{\bibinfo{volume}{79}},
  \bibinfo{pages}{3238} (\bibinfo{year}{1997}).

\bibitem[{\citenamefont{Grigoriev}(2001)}]{2DLKth1}
\bibinfo{author}{\bibfnamefont{P.}~\bibnamefont{Grigoriev}},
  \bibinfo{journal}{Sov. Phys. JETP} \textbf{\bibinfo{volume}{92}},
  \bibinfo{pages}{1090} (\bibinfo{year}{2001}).

\bibitem[{\citenamefont{Itskovsky and Maniv}(2001)}]{2DLKth2}
\bibinfo{author}{\bibfnamefont{M.~A.} \bibnamefont{Itskovsky}}
  \bibnamefont{and} \bibinfo{author}{\bibfnamefont{T.}~\bibnamefont{Maniv}},
  \bibinfo{journal}{Phys. Rev. B} \textbf{\bibinfo{volume}{64}},
  \bibinfo{pages}{174421} (\bibinfo{year}{2001}).

\bibitem[{\citenamefont{Martin et~al.}(2003)\citenamefont{Martin, Maslov, and
  Reizer}}]{Maslov}
\bibinfo{author}{\bibfnamefont{G.~W.} \bibnamefont{Martin}},
  \bibinfo{author}{\bibfnamefont{D.~L.} \bibnamefont{Maslov}},
  \bibnamefont{and} \bibinfo{author}{\bibfnamefont{M.~Y.}
  \bibnamefont{Reizer}}, \bibinfo{journal}{Phys. Rev. B}
  \textbf{\bibinfo{volume}{68}}, \bibinfo{pages}{241309(R)}
  (\bibinfo{year}{2003}).

\bibitem[{\citenamefont{Syed et~al.}(2004)}]{inhomogeneity}
\bibinfo{author}{\bibfnamefont{S.}~\bibnamefont{Syed}} \bibnamefont{et~al.},
  \bibinfo{journal}{Appl. Phys. Lett.} \textbf{\bibinfo{volume}{84}},
  \bibinfo{pages}{1507} (\bibinfo{year}{2004}).

\bibitem[{\citenamefont{Ballestad et~al.}(2001)}]{MBEGaAs}
\bibinfo{author}{\bibfnamefont{A.}~\bibnamefont{Ballestad}}
  \bibnamefont{et~al.}, \bibinfo{journal}{Phys. Rev. Lett.}
  \textbf{\bibinfo{volume}{86}}, \bibinfo{pages}{2377} (\bibinfo{year}{2001}).

\bibitem[{\citenamefont{DasSarma and Mason}(1985)}]{nonparabolicity}
\bibinfo{author}{\bibfnamefont{S.}~\bibnamefont{Das Sarma}} \bibnamefont{and}
  \bibinfo{author}{\bibfnamefont{B.~A.} \bibnamefont{Mason}},
  \bibinfo{journal}{Phys. Rev. B} \textbf{\bibinfo{volume}{31}},
  \bibinfo{pages}{R1177} (\bibinfo{year}{1985}).

\bibitem[{\citenamefont{Kwon et~al.}(1994)\citenamefont{Kwon, Ceperley, and
  Martin}}]{KCM}
\bibinfo{author}{\bibfnamefont{Y.}~\bibnamefont{Kwon}},
  \bibinfo{author}{\bibfnamefont{D.~M.} \bibnamefont{Ceperley}},
  \bibnamefont{and} \bibinfo{author}{\bibfnamefont{R.~M.} \bibnamefont{Martin}},
  \bibinfo{journal}{Phys. Rev. B} \textbf{\bibinfo{volume}{50}},
  \bibinfo{pages}{1684} (\bibinfo{year}{1994}).


\bibitem[{\citenamefont{Asgari et~al.}(2004)}]{Polini}
\bibinfo{author}{\bibfnamefont{R.}~\bibnamefont{Asgari}} \bibnamefont{et~al.}
  (\bibinfo{year}{2004}), \eprint{cond-mat/0406676}.

\end{thebibliography}

\end{document}